\documentclass[]{elsarticle}
\usepackage{amsmath}
\usepackage{array}
\usepackage{makecell}
    \usepackage{tikz}

  \def\A{\mathcal{A}}

\def\E{\mathcal{E}}

\def\G{\mathcal{G}}
\newtheorem{lemma}{Lemma}[section]
\newtheorem{proposition}[lemma]{Proposition}

\title{The existence and abundance of ghost ancestors in biparental populations}

\author{Simon Gravel\fnref{label2} }

\ead{simon.gravel@mcGill.ca}
\address[label2]{Genome Quebec Innovation Centre\\ Department of Human Genetics,
 McGill University\\740 Dr Penfield\\ Montreal, Quebec\\ H3A 0G1, Canada;\\simon.gravel@mcgill.ca}
 
 \author{Mike Steel\fnref{label3} }
 \address[label3]{Biomathematics Research Centre, University of Canterbury, New Zealand}

\begin{document}
\begin{frontmatter}

\begin{abstract}
In a randomly-mating biparental population of size $N$ there are, with high probability, individuals who
are genealogical ancestors of every extant individual within approximately $\log_2(N)$ generations into the past. We use this result of J. Chang to prove a curious corollary under standard models of recombination: there  exist, with high probability, individuals within a constant multiple of $ \log_2(N)$ generations into the past
who are simultaneously (i) genealogical ancestors of {\em each} of the individuals at the present, and (ii) genetic ancestors to {\em none} of the individuals at the present.  Such ancestral individuals -  ancestors of everyone today that left no genetic trace --  represent  `ghost' ancestors in a strong sense.  In this short note, we use simple analytical argument and  simulations to estimate how many such individuals exist in finite Wright-Fisher populations. 
\end{abstract}

\end{frontmatter}

\section{Introduction}

The reproductive success of an individual is traditionally measured by the number of offspring it produces.  However, even given a fixed number of descendants, there is a fair amount of variation in the amount of genetic material left by an individual.  An ancestor that contributes genetic material to an individual through multiple lineages is expected to leave more genetic material than an ancestor that contributes along a single lineage, for example \citep{ Baird, bar,   lac,  man, mat}. Even if we fix the genealogy, there remains considerable stochasticity in the amount of genetic material left by each ancestor during meiosis. The probability of inheriting genetic material along a given ancestral lineage decreases rapidly with the number of generations. Thus, an individual leaving a single descendant after more than 10 generations is likely to leave no genealogical material in future generations, even if this descendant is wildly prolific. This limits the information about past genealogies that can be recovered from present-day genomic data \cite{tha, tha2}.  Not only are we limited to individuals that left descendants, but even the genomes of individuals that left offspring may be completely inaccessible.  

A result of Chang \citep{cha, don2} states that first-generation (or \emph{founding}) individuals of a constant-sized, randomly mating Wright-Fisher diploid population rapidly reach one of two ancestral states: about $80\%$ become ancestral to the complete subsequent population, and $20\%$ leave no descendants. We are interested in the $80\%$ of `successful' individuals, and wonder about the proportion of such individuals that leave no genetic material to any of their large number of descendants. 
 The question of whether genealogical ancestors leave genetic material in an infinite population was discussed in detail in \cite{Baird}, where a branching approximation to the infinite-population Wright-Fisher model was used to show a logarithmic decrease of survival probability with time. Here we consider the finite population case and obtain analytical and numerical results under the exact Wright-Fisher model. Because the approximations of \cite{Baird} are reasonable, we expect that our results should agree in the large-$N$ and long-time limits, but our results will provide additional insight for small population and short times.      

 We first formally show that, amongst those individuals who are genealogically ancestral to the complete population, the proportion that leaves no genetic material at the present is nonzero in a finite population, and indeed approaches 1 when the number of generations and population size become large. We turn to simulation to estimate their abundance for finite genomes and finite number of generations. We find that the proportion of ghost ancestors grows approximately logarithmically with the number of generations in a constant-size population and that convergence to the large-population limits occurs rapidly at short time-scales. 
 
Our work complements recent efforts to study the relationship between genealogical and genetic ancestries. Wakeley et al. \cite{wak} recently proposed an improved approach to model the recent pedigree structure of samples within coalescent theory, but found that this improved modeling had a relatively modest effect on the simulated statistics. Wiuf and Hein \cite{wiu} used mathematical modelling and simulations to study the distribution of ancestral material of an extant chromosome, with the aim of addressing two questions: (i) how many ancestors are there to a present human chromosome?  and (ii) how many different sequences in an ancestral population can one sample by sequencing extant sequences?
More recently, Matsen and Evans \cite{mat} used simulations and probabilistic analysis to investigate relationships between the number of descendant alleles of an ancestor allele and the number of genealogical descendants of the individual who possessed that allele. Here we explore the impact of demography on the short- and long-term probability that an individual with many genealogical descendants leaves no genetic material. We provide a short formal proof for the existence of such individuals, in a probabilistic sense, and study their prevalence under different demographic scenarios with Wright-Fisher reproduction.

\section{Ghost and super-ghost ancestors}

Given a population at the present, an ancestral individual $I$ is said to be a {\em ghost} ancestor if:
\begin{itemize}
\item[(i)]  $I$ is the genealogical ancestor of at least one individual at the present, and
\item[(ii)] $I$ contributes nothing genetically to any individual at the present.
\end{itemize}
A {\em super-ghost} replaces condition (i) with the stronger condition:

\begin{itemize}
\item[(i)$'$]  $I$ is the genealogical ancestor of {\em all} individuals at the present.
\end{itemize}

\begin{figure}
\begin{center}
\scalebox{0.4}{ \includegraphics{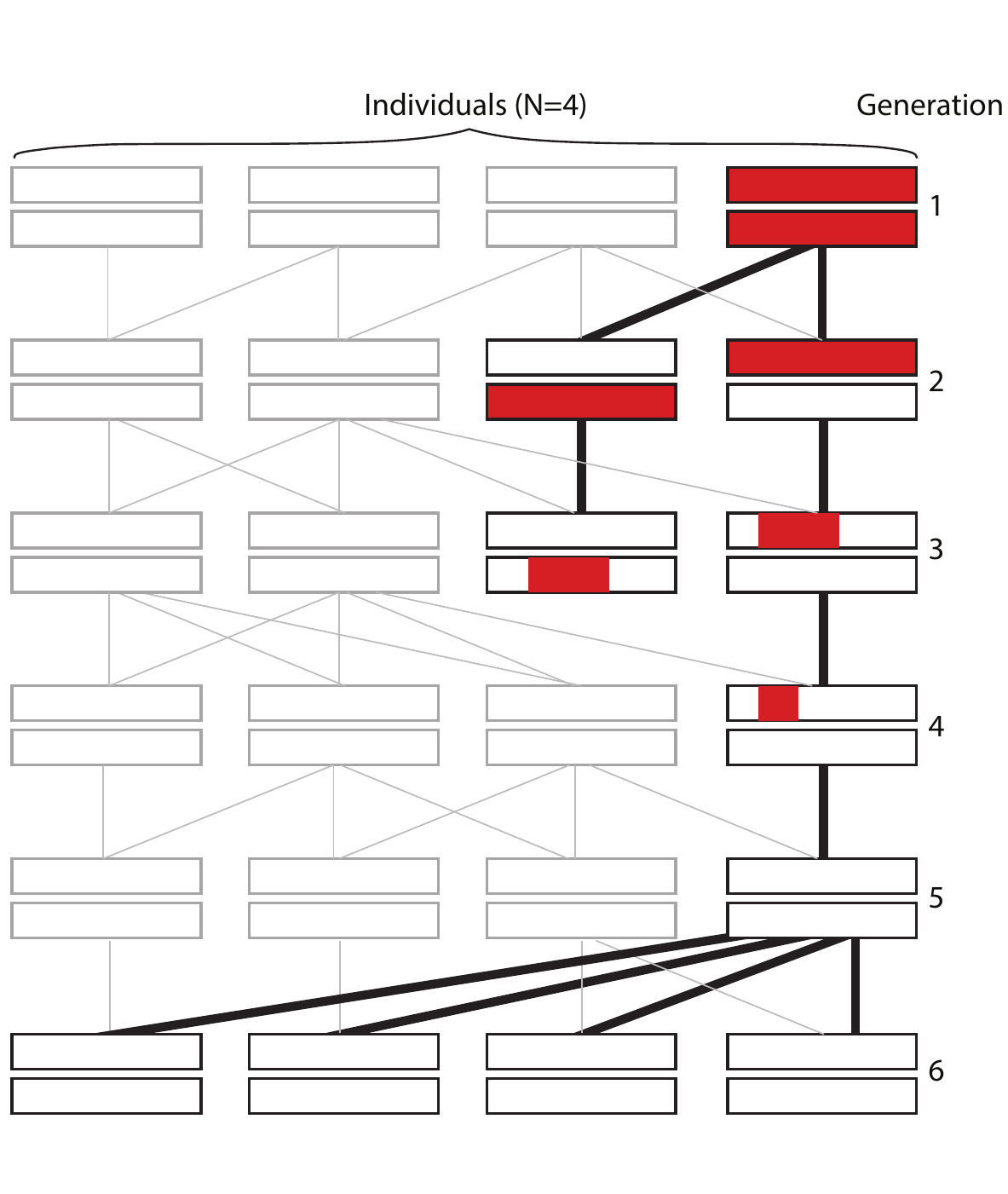} }
 \caption{\label{sup}  In this schematic example, the rightmost individual in generation $1$ has genealogical descendants until generation $6,$ but left no genetic material in generations $5$ and $6$. By generation $5$ it has become a ghost; by generation $6$ it is a `super-ghost',  in that it is now ancestral to the entire population.}
 \end{center}
 \end{figure}

A schematic example of a super-ghost is shown in Fig.~\ref{sup}.

\subsection{The existence of super-ancestors: Chang's result}

Consider the genealogical ancestry of a randomly-mating biparental population under the neutral Wright-Fisher model. In this model,  generations are 
discrete, and each individual selects two parents uniformly at random from the previous generation, and this process is extended back in time independently from generation to generation. This generates a random genealogy for the population. 

Because each individual has on average two offspring per generation, we would expect individuals to leave a large number of offspring if their descending lineage survives the stochasticity of the first few generations. In fact, the probability of eventually becoming ancestral to the \emph{complete} population in a randomly mating population of size $N$ converges to $1-\rho\approx 0.7968$ for large $N$, where $\rho$ satisfies $\rho = e^{-2+2 \rho}$  \cite{cha}  (see also  \cite{der1, der2, der3}).    
Because the expected number of descendants of successful individuals initially grows close to exponentially, we would expect that it must take at least  $\log_2(N)$ generations to become ancestral to the complete population. In fact, Chang  \cite{cha} established the following two tight asymptotic results that are key to this paper: 
\begin{itemize}
\item[(a)] the  number of generations $G_1$ required to find the first super-ancestor approaches $\log_2(N)$, in the sense that  $\frac{G_1}{\log_2(N)}$ converges in probability to 1 as $N$ grows;  and 
\item[(b)] the number of generations $G_a$ until \emph{all} individuals that left descendants are super-ancestors approaches $(1+\zeta) \log_2(N),$ where $\zeta \approx 0.7698$, in the sense that $\frac{G_a}{(1+\zeta)  \log_2(N)}$  converges in probability to 1 as $N$ grows. 
\end{itemize}

%

\subsection{The existence of super-ghosts}

We consider again the genealogical model from \cite{cha}, and model the transmission of genetic material. Given a genealogy, each individual receives one copy of each chromosome from each parent. An individual transmits chromosomes by recombining the two chromosome copies it has inherited. We model recombination as a Markovian copying process along the chromosome length: starting from one edge, the transmitted chromosome is generated by copying one of the two inherited chromosomal copies, with a transition rate between the two of 1 per Morgan. 

\begin{proposition}
\label{pro-first}
Under the random biparental genealogical model with discrete generations of size $N$, the probability that there is a super-ghost ancestor within $c \cdot \log_2(N)$ generations from the present converges to 1 as $N$ grows, for any $c > 1$.
\end{proposition}

The formal proof of this proposition is provided in the Appendix, and here we provide an informal argument for the existence of super-ghosts and logarithmic dependence on $N$.  The proof has two steps: first, we show that ghosts exist, i.e., that there are individuals that leave genealogical descendants but no genetic material. Second, we show that the descendants are likely to become ancestral to the entire population, ensuring that the ancestral individual is now a super-ghost. To estimate the timing it is convenient to proceed back in time.
In a Wright-Fisher model, the probability that an individual is the unique descendant to an individual living $G$ generations is nonzero. It can be easily calculated through dynamic programming, and is trivially bounded below by $P(1)^G, $ where $P(1)$ is the probability of being an only child. In this particular case of a single descending lineage, we can also easily show that the probability of leaving no genetic material becomes nonzero in any reasonable model of recombination: The descendant haploid genome received genetic contributions along $2^G$ lineages (some of the lineages may lead to the same ancestors). However, the expected number of inherited recombination breakpoints that occurred in the intervening generations is simply $GL$, where $L$ is the length of the genome, in Morgans. The expected number of independently inherited blocks is therefore $GL+C,$ where $C$ is the number of chromosomes. If the number of lineages is greater than the number of blocks ($2^G>GL+C$), there must be some lineages that have not contributed genetic material. Since the probability of having received genetic material is independent of the lineage in the Wright-Fisher model, we must have a finite probability that our ancestor did not leave genetic material to the single descendant after $G$ generations. Finally, using Chang's result, this descendant has a probability of about $1-\rho\approx 0.7968$ of becoming the ancestor of the complete population within $(1+\zeta) \log_2(N)$ generations. The formal proof explains how this can be extended to any $c>1.$

\bigskip

To estimate the prevalence of super-ghosts after a long time has elapsed, consider again the diploid Wright-Fisher model with discrete generations of size $N$, and a model of random genetic recombination involving genomes of $B$ base pairs.  For a fixed population size $N$, let 
\begin{itemize}
\item $q(N,T)$ be the probability that a randomly selected individual $T$ generations ago contributes nothing genetically to the present generation;
\item $q'(N,T)$ be the probability that a randomly selected individual $T$ generations ago is a super-ghost.
\end{itemize}

\begin{proposition}
\label{pro-second}
\mbox{ } 
\begin{itemize}
\item[{\rm (i)}]
$\lim_{N \rightarrow \infty} \lim_{T \rightarrow \infty} q(N,T) = 1,$
\item[{\rm (ii)}] $\lim_{N \rightarrow \infty} \lim_{T \rightarrow \infty} q'(N,T) = 1-\rho\approx 0.7968.$
\end{itemize}
\end{proposition}

Once again, the  formal proof of this proposition is provided in the Appendix, but the informal argument is straightforward: in a finite, constant-size population, genetic drift at each site will eventually lead to the fixation in the population of the allele inherited from a single ancestor. If the genome is finite, we only have to wait a finite amount of time until each site has fixed, and at most $B$ ancestors will prevail. If we increase $N$ while leaving $B$ fixed, the proportion of individuals who contribute genetic material can be made arbitrarily small. Note that this proof of Proposition \ref{pro-second} requires a genome with a finite number $B$ of base pairs, contrary to other results in this article, which also hold in the continuous limit where $B$ goes to infinity as the length of the genome $L$ is held constant. It may be possible to generalize the proof to a continuous genome, however we were not able to produce such a proof.  

The order of the limits does matter in Proposition \eqref{pro-second}, since taking the large-$N$ limit first would make it impossible for an individual to become ancestral to the complete population. There are many ways to relax our definition of `successful' to make it realizable in this context (e.g., leaving a finite but large number of descendants), and we conjecture that in that context,  the majority of genealogically `successful' ancestors leave no genetic material independent of the limit order. We expect that the convergence to the large-$N$ limit is rapid for short time-frames: the randomness of individual genealogies is the dominant stochastic factor, and is relatively unaffected by the finite population size.

\section{Simulations}

We performed diploid Wright-Fisher simulations, keeping track of the list of ancestors for each individual and the ancestral contributor of each inherited genetic segment.  We simulated 36 chromosomes of length 1 Morgan to mimic the length of the human genome, while preserving a convenient chromosome exchangeability. We simulated populations of sizes 20 to 4000 over 800 generations. Each population was run multiple times, such that we had simulated the descendants of 20,000 ancestors for each population size. 

\begin{figure}
\scalebox{0.8}{ \includegraphics{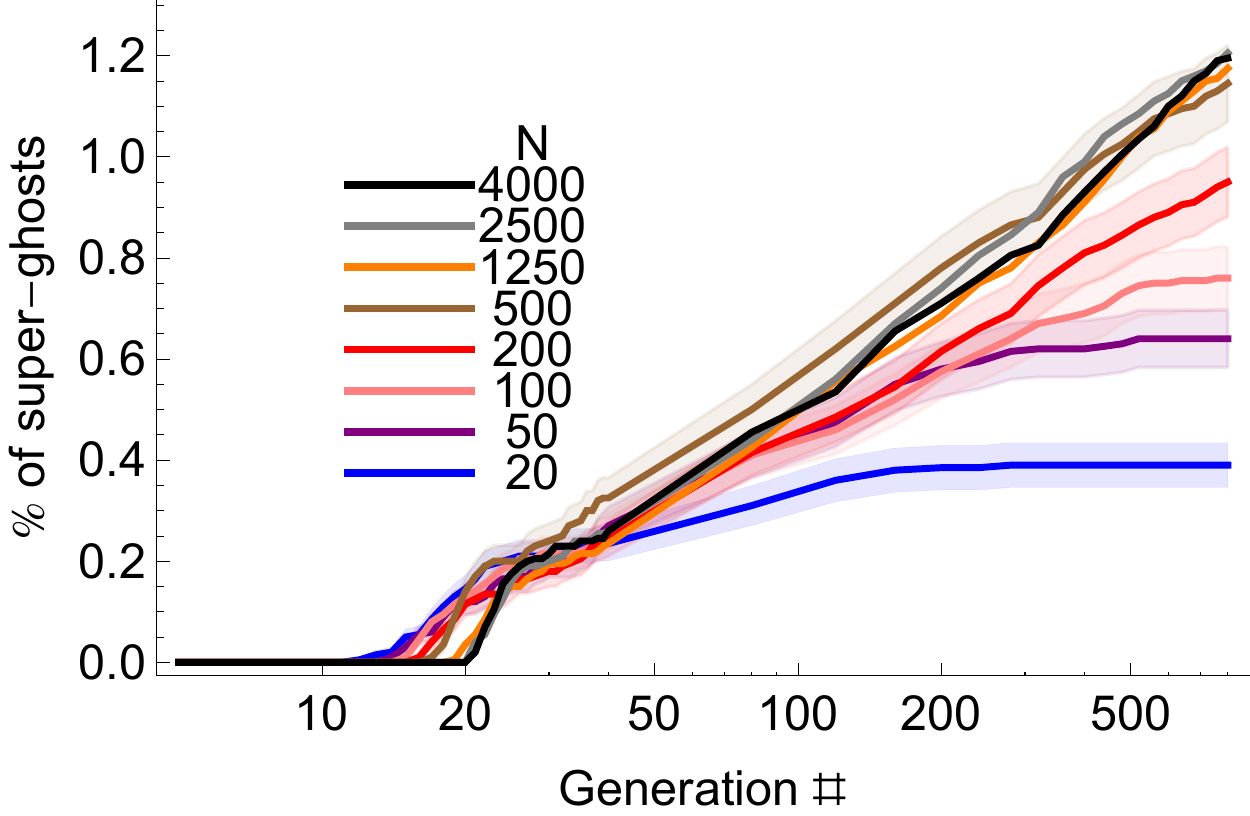} }
 \caption{\label{gvsN} Probability that a founding individual is a super-ghost as a function of the time elapsed, for different population sizes.  Independent replicates were performed for each population size so that 20,000 individuals were simulated for each population size. Shaded areas represent 1-sigma deviations assuming that the number of ghosts is Poisson distributed.}
 \end{figure}

The earliest occurrence of a super-ghost was after 12 generations in a population of size 20, and after 21 generations in a population of 4000 (Fig. \ref{gvsN}). This variation is due to the slightly longer time needed to become ancestral for the whole population in the larger population, as expected from Chang's result. Soon after, all individuals are either genealogical ancestors to the whole population, or to no one. 
 
The initial rate of super-ghost creation in that stage becomes roughly independent of population size, and the number of super-ghosts, once they first appear, grows approximately logarithmically with $T$ for $N>100$ . As argued above,  this initial independence on population size makes sense; since alleles inherited from ghost ancestors are ultimately unsuccessful, they are unlikely to have reached frequencies high enough that the finite population matters. Eventually, some alleles reach sufficient frequency to be affected by the finite population size. Some of them will fix in the population, guaranteeing that the ancestral individual will never become a ghost. In this way, drift reduces the probability of becoming a ghost.

\begin{figure}
\scalebox{0.5}{
\includegraphics{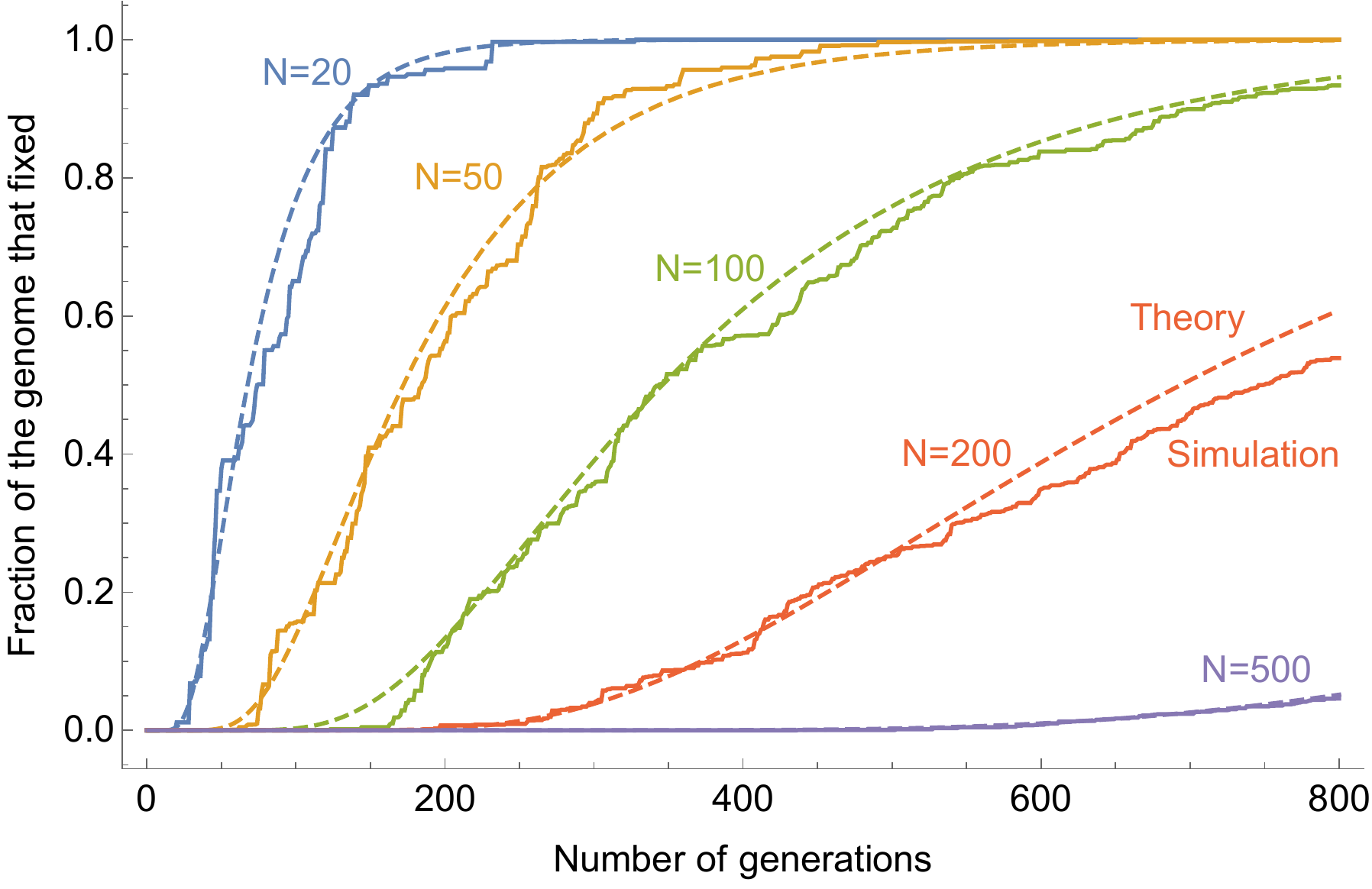}}
\scalebox{0.7}{\includegraphics{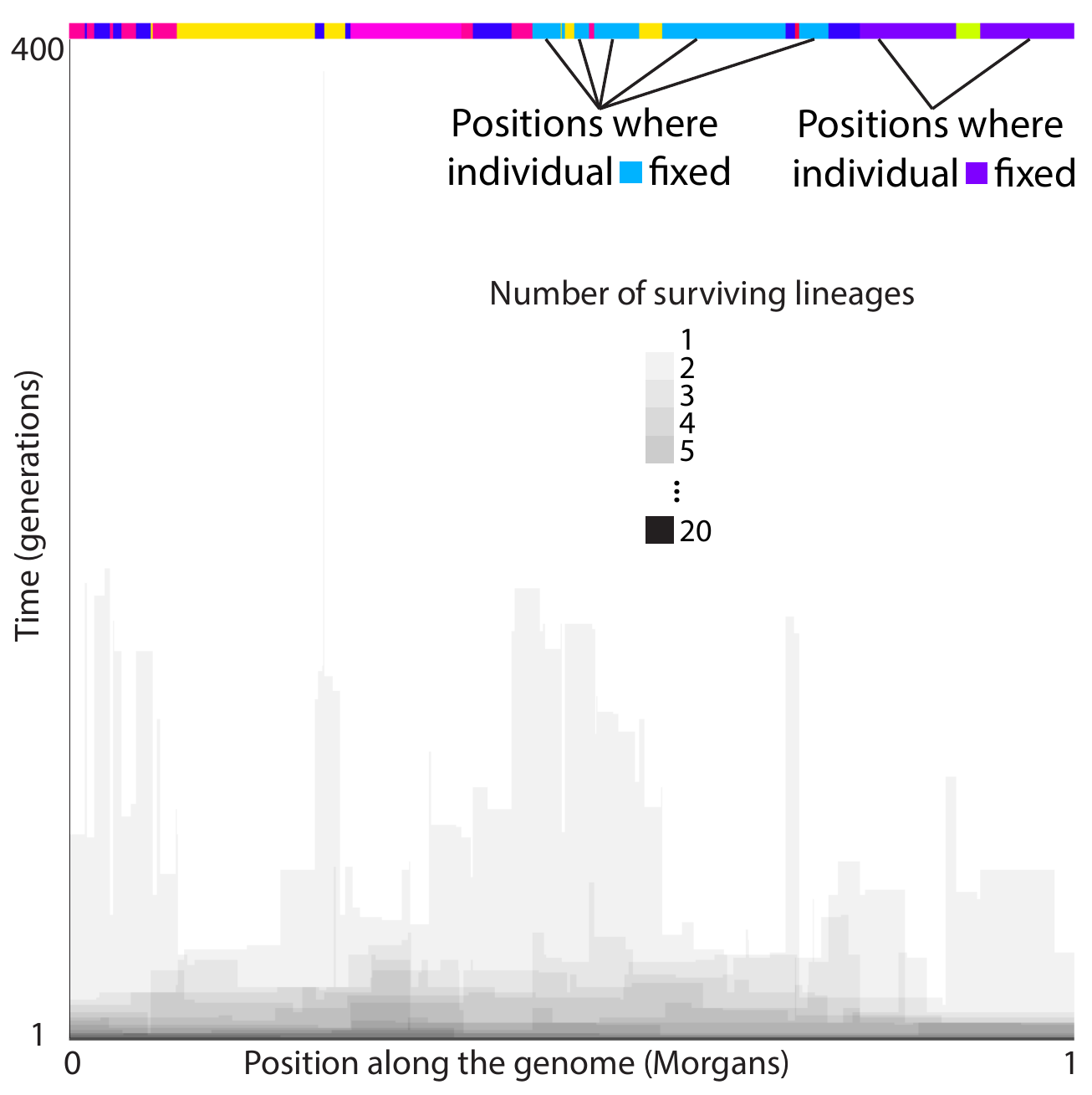}}
\caption{\label{linkagemap} (Top) Proportion of the genome fixed for any ancestral haplotype. Simulated values are taken from a single simulated population by population size. Theoretical estimates are derived from Equation 3.27 in \cite{wakeley}. (Bottom) Number of individuals who left descendants at each position along a chromosome of length $1$ Morgan in a population of size 20. The colored row shows the identity of the surviving individual at each position---there are 8 surviving lineages in this example.}
\end{figure}

The number of ancestral individuals who are ghosts should stop increasing once each site along the genome is fixed for a given ancestral haplotype, since individuals whose haplotypes fixed in the population are guaranteed to leave genetic descendants. We can observe this for the smaller populations on Fig. \ref{gvsN}. Fig. \ref{linkagemap} shows how the patterns of fixation vary with population size, and how they correlate along the genome. 

  For the population of size 20, only $0.4\%$ of ancestors ended up as super-ghosts, much less than one per simulated population.  The proportion reaches $1.2\%$ in the populations larger than $1000$ after $800$ generations. Simulations up to 2500 generations in populations of size $2500$ showed a continuation of the logarithmic increase, with $1.8\%$ of super-ghosts at that point. Assuming a constant population size of $10,000$ and a generation time of $30y$, the number of super-ghosts living between 10,000 and 50,000 years ago would therefore be a bit above 100, in that model. However, population structure and population size changes are likely to affect the proportion of ghost ancestors.

\begin{figure}
\scalebox{0.8}{ \includegraphics{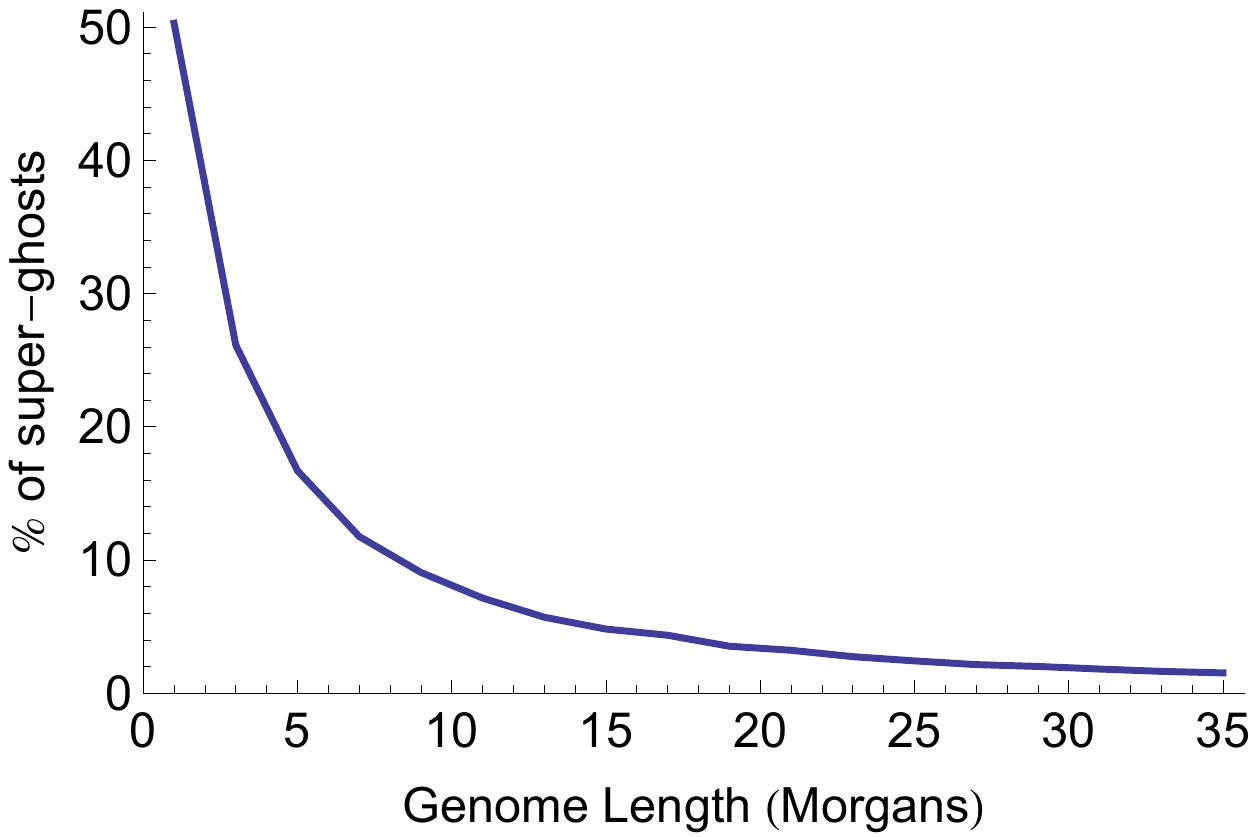} }
 \caption{\label{gvsL} Probability that a founding individual is a super-ghost as a function of genome length, after 2500 generations in a population of size 2500.  Three replicate simulations were performed with 36 unit-length chromosomes, and results for each genome length are obtained by downsampling the chromosomes 10 times and averaging the results.  }
 \end{figure}

Simulations also provide insight into the different factors that influence the proportion of ghosts. For example, we expect that the probability of being a ghost decreases with the length of the genome, since a longer genome means more opportunities to inherit genetic material. The proportion of super-ghosts in the population as a function of  genome length, estimated from simulations, is shown in Fig. \ref{gvsL}.  The long tail observed in this graph indicates that the probability of leaving genetic material along different chromosomes is strongly correlated. Because of the independent assortment of chromosomes, these correlations can only be mediated by the shared genealogies: individuals that left no genetic material over the first 25 chromosomes are likely to have genealogies favorable to becoming ghosts. 

\begin{figure}
\scalebox{.8}{\includegraphics{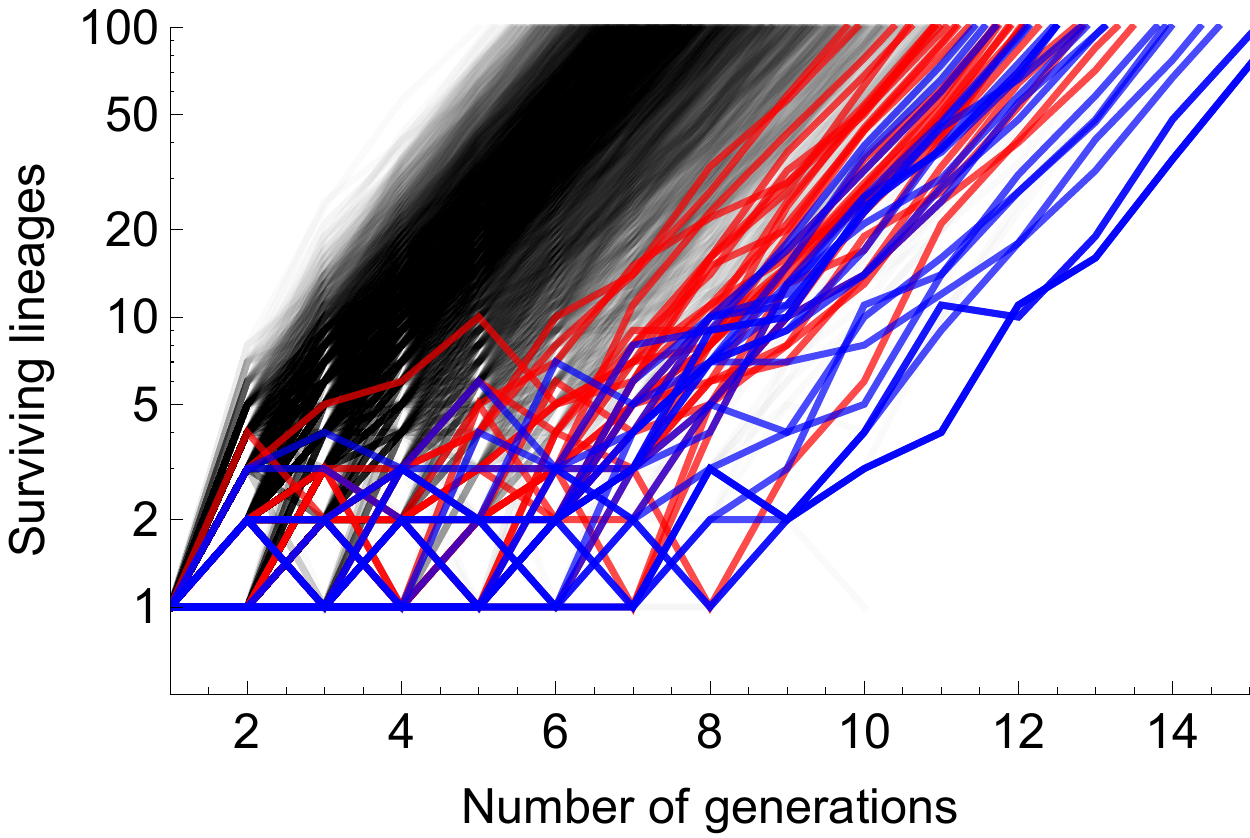}}
\caption{\label{earlytraj}Number of offspring per generation per individual in a diploid Wright-Fisher population of 2500 individuals over 800 generations. Individuals that will become super-ghosts are colored according to the number of generations that were necessary to achieve ghost status--blue for generations 1 to 150, red for 151-800. 
}
\end{figure}

Such ghost-friendly genealogies occur if a single descendant survives after many generations, because the probability of inheriting genetic material along a single lineage decreases rapidly with the number of generations, and the probability that a lone descendant is an ancestor to the complete future population is close to $80\%$. We find that in fact all super-ghosts follow a trajectory with few offspring over many initial generations  and are in the tail of the distribution with regard to this feature (see Fig. \ref{earlytraj}), with a tendency for the more extreme genealogies to lead to earlier ghosts status. Genealogies that maintain a small number of offspring for many generations are rare because the expected initial growth rate in the number of offspring is 2 in a constant-size diploid population; the number of offspring either reaches 0 or increases rapidly. 

Based on this intuition, we expect that the number of individuals that will become ghosts is sensitive to what happens in the first few generations. To investigate this, we consider a few demographic scenarios: exponential growth and exponential decline ($r=\pm 0.2\%$), early short bottlenecks, and early population spikes. Even though the models of population growth and decline lead to 5-fold population size changes over the 800 generations of the simulation, the impact on the final number of ghosts is at most modest for populations that have reached fixation. The number of super ghosts in $1000$ simulated populations of size $20$ over $800$ generations was $78$ in the constant-size population ($0.39\%$ of individuals), $68$ in the declining population ($0.34\%$  of individuals), and $76$ in the growth model ($0.38\%$ of individuals). The difference was more important for larger populations where fixation had not yet been reached:  in the population of size $500$, the number of super ghosts was $229$ in the constant-size population, compared to $375$ in the declining population. We can interpret the difference as a combination of two factors: first, the rate of genetic drift accelerates over time in a decreasing population, such that the initial rate of ghost creation is expected to be elevated compared to a population whose size does not decrease. Second, even with equal drift, fixation can be reached earlier in a smaller population because the available frequencies are more coarsely discretized. If we let the declining population evolve for $478$ generations, to ensure an equal amount of drift to the constant-size population, the number of super ghosts is reduced to $274$, and the remaining difference with the constant-size population is likely due to the earlier fixation. These two effects can substantially affect the initial rate of ghost creation, but do not necessarily lead to an increase in the long-time ghost probability, as we saw for the smaller populations. 

\begin{figure}
\scalebox{.8}{\includegraphics{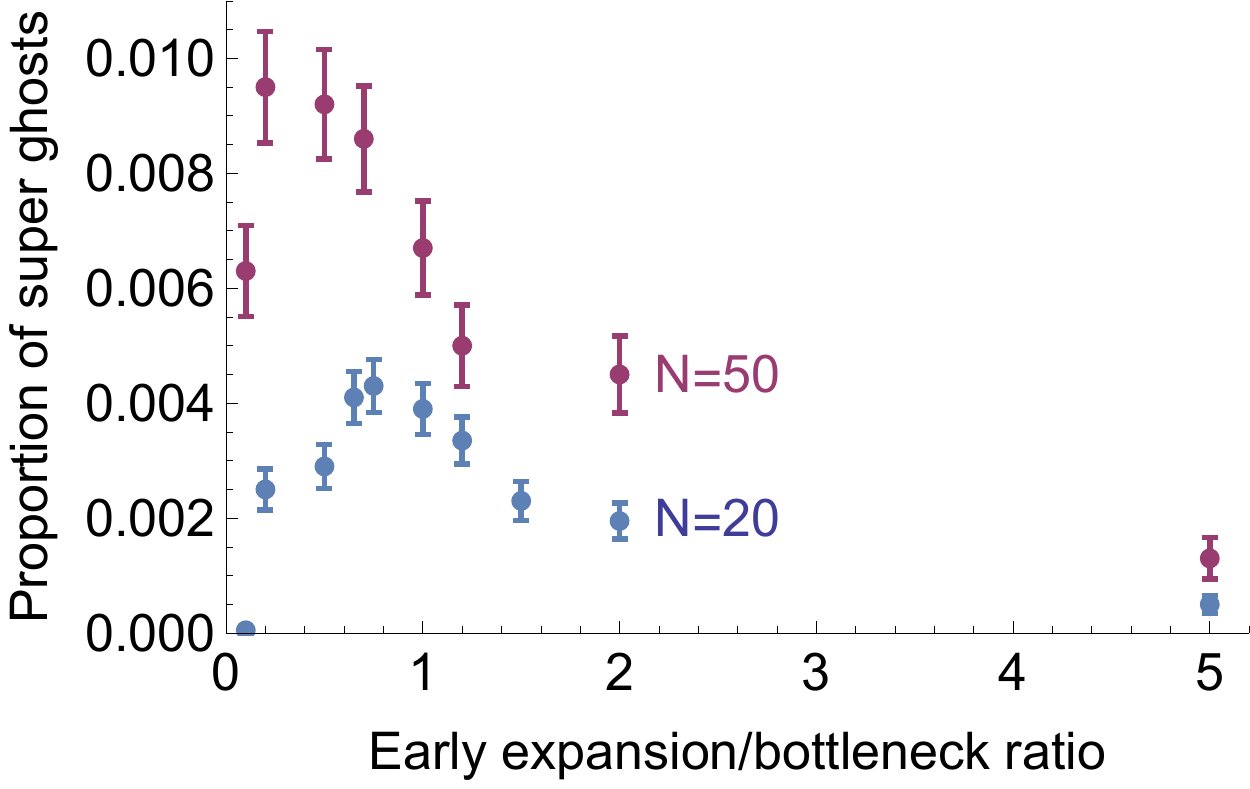}}
\scalebox{.8}{\includegraphics{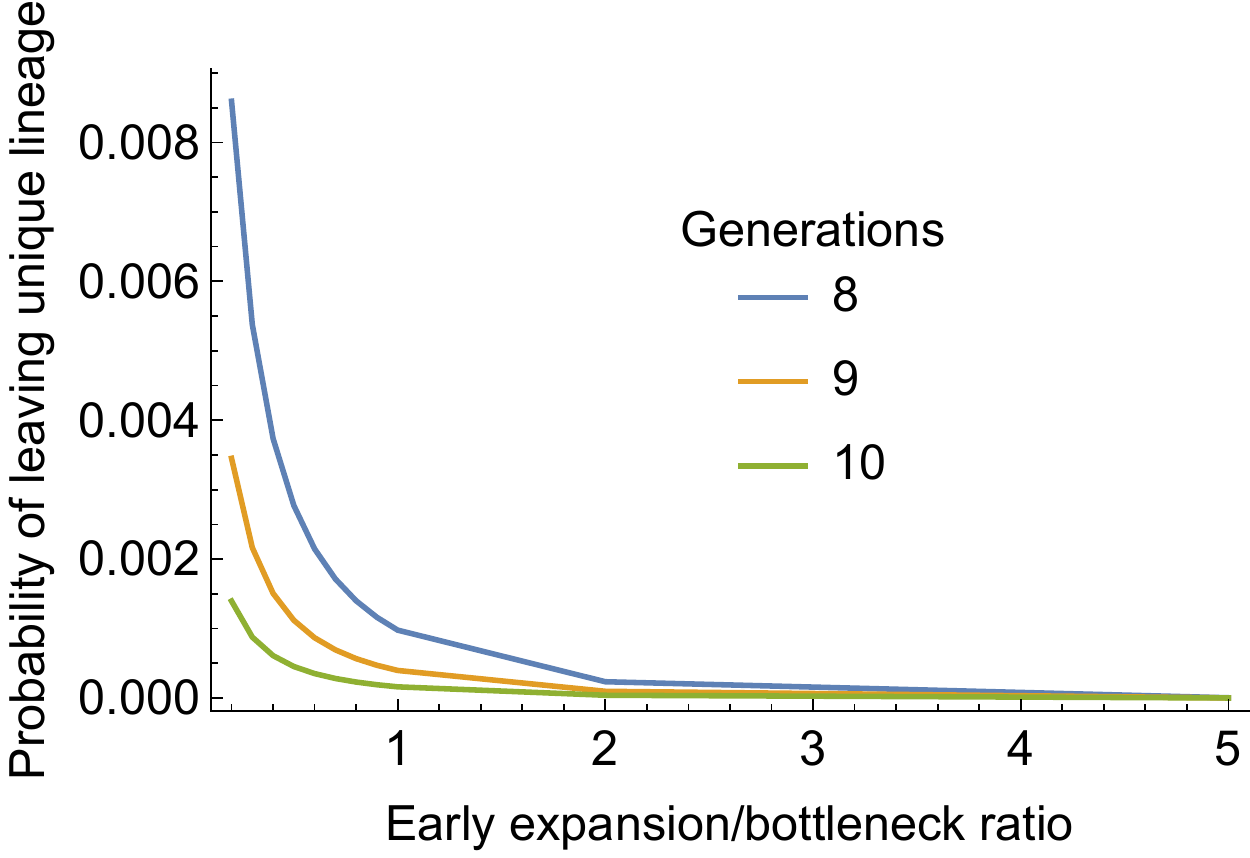}}
\caption{\label{bumps} (Top) Probability of becoming a super ghost for different perturbations of the population size. Populations of size 20 and 50 were simulated for 800 generations. Between generations 5 and 15, the population size was scaled by factors ranging between 0.1 and 5 to simulate early bottleneck and expansions. (Bottom) Probability that an individual leaves exactly one descendant after 8, 9, or 10 generations for an infinite population.   
}
\end{figure}

To study the effect of rapid population size changes in the early population history, we considered a piece-wise population change between generation $5$ and generation $15$, with results displayed in Fig. \ref{bumps}. We find that the probability of becoming a ghost reaches a maximum for decreases in population size. Early expansions dramatically reduce the creation of super-ghosts, whereas bottlenecks can increase the proportion of ghosts (Figure \ref{bumps}, Top). The intuition for this behavior is presented in Figure \ref{bumps}, bottom. By generation $5$, ancestors have an expected $2^5=32$ descending lineages. In a large population, these lineages correspond to distinct individuals, so that an abrupt population decline vastly increases the probability of leaving a small number of descendants. In smaller populations, the descendants are less likely to be distinct, and the abrupt decline makes it more likely that an ancestor leaves no descendants, and increases the likelihood that survivors fix in the population. When the population returns to its initial size at generation $15$, most individuals left either $0$ or a large number of descendants, so population size fluctuations are unlikely to create single-descendant lineages. 
   
\section{Discussion}

We find approximately logarithmic growth of the probability of becoming a ghost starting between $10$ and $20$ generations (that is, we see linear behavior on the log-scale, Fig. \ref{gvsN}). This is reminiscent of the prediction \cite{Baird} that a haploid block of Morgan length $y<1$ would leave descendants with probability $\frac{y}{\log(y G)}$ after $G$ generations for $\log(yG)>>1$. Thus, both short- and long- time dynamics seem to depend on $\log(G)$. This logarithmic slow-down explains why we may, at the same time, lose all genetic information about relatively recent ancestors while retaining some information about many very distant ancestors. According to simulations, approximately $0.5\%$ of individuals are expected to be super-ghosts in a large, constant-size population after 100 generationsm, and it reaches $1.4\%$ after 1000 generations. These results are somewhat higher than those of  \cite{Baird} et al, who found $1.9\%$ survival rate for a haploid genome after 1000 generations, which corresponds to approximately $0.77\%$ in a diploid genome (see Appendix). The difference is likely due to the limited sample size of $100$ early genealogies used in \cite{Baird} for this particular calculation.  A key feature of individual histories that determines persistence in the population is the number of surviving lineages in the first few generations after an individual lived. The number of ``favorable'' genealogies -- falling in the lower right of Fig.  \ref{earlytraj} -- may be as low as 2 in the  analysis of \cite{Baird}, causing a high relative variance.  

Demography affects the probability of becoming a ghost and the speed at which this happens. Convergence to the infinite-population limit appears accurate for $N>G$, a regime where the effect of drift is relatively weak. For small populations, the rate of creation of super-ghosts decreases for $G>N$, approaching a plateau: each individual ancestral allele has a finite probability of fixing in the complete population, ensuring indefinite descendance. Once all individual genetic contributions have either fixed or disappeared, there can be no more new ghosts.  The probability that an individual contributes genetically to descendants at infinite time depends on the slow race between drift, which fixes or eliminates haplotypes in the population, and recombination, which creates new haplotypes at unfixed sites. This race will eventually be interrupted by the discreteness of the genome, leading to the result from Proposition \ref{pro-second} that the probability of leaving genetic material goes to zero as the population size goes to infinity. The same is true in the continuous genome limit in the model of \cite{Baird}. However, the approach to the limit where all individuals become ghosts may be slow for both time $G$ and population size $N$. 

The probability of leaving genetic material in finite populations is unknown beyond the cases simulated here, and we lack even approximate expressions. However, our simulations suggest relatively rapid convergence to the infinite-population-size limit, and we may venture a few additional conjectures: A) the probability of leaving genetic material decreases monotonically with population size, as suggested in Fig. \ref{gvsN}. B) the majority of the eventual fixation events occur within the first $c N$ generations for some sufficiently large constant $c$ (as suggested by Fig. \ref{linkagemap} ; C) the infinite-population size estimate is accurate for $G<N$ (as suggested in  Fig. \ref{gvsN}).
If these hold, the long-term probability of leaving genetic material in a population of size $N$ lies between the infinite-population results at times $N$ and $cN$. Because of the extremely slow rate of ghost creation, the resulting interval in ghost probability may be quite narrow. Extrapolating the observed logarithmic growth and assuming $c=1000$, the survival probability at \emph{infinite} time of an individual in a population of size $7$ billion would be between $69\%$ and $72\%$ in the continuous-genome limit. 

However, such predictions are difficult to apply to actual populations, because of the potential  large effect that early departures from model assumptions can have on the long-term dynamics of the system. We found that even very short bottlenecks and population size increases can have a large impact on genetic survival probability. We therefore expect that even slight departures from the Wright-Fisher reproduction can have similar large effects on survival probability. In the long term, population structure and migrations can dramatically affect the rate at which genetic material fixes and disappears.  We found that short-lived, early population size fluctuations had a large impact on the long-term genetic contributions of individuals, with bottlenecks leading to the highest probability of becoming a super ghost. 

These results emphasize that early genealogies can have a massive impact on the long-term dynamics of a population when statistical outliers are important. Modelling of pedigree structure, including effects such as non-Poisson offspring distribution, correlation in offspring number among successive generations, and range expansions ({\em c.f.} \cite{Mor}) are therefore key to a better understanding of long-term evolutionary outcomes.

\section{Acknowledgements} We thank two anonymous reviewers, and the handling editor for their suggestions on an earlier version of this paper. We also thank Armandine Veber for two helpful comments concerning the technical aspects of the proofs. MS thanks the Allan Wilson Centre for funding support. SG thanks CIHR operating grant MOP-134855 and the CIHR Canada Research Chair in Statistical and Population Genetics.

\section{Appendix}

{\em Proof of Proposition~\ref{pro-first}}

Under a  biparental model, for any individual $v$ we will say that  an ancestor $w$ of $v$ from $k$ generations earlier is a {\em $k$-critical ancestor  for $v$} if
 the following two properties hold:
\begin{itemize}
\item[(i)] $w$ leaves no genetic contribution in $v$;  and 
\item[(ii)] the only genealogical descendant of $w$ in the same generation as $v$ is $v$.  
\end{itemize}

If a common ancestor of all individuals at the present has a $k$-critical ancestor $w$, then $w$ is an ancestral super-ghost. We will attempt to produce such an ancestor, and our first step is to produce a $k$-critical ancestor.

Let $p_k(v)$ be the probability that $v$ has a $k$-critical ancestor.  Under the random biparental genealogical  model, $p_k(v)$ is the same for all individuals $v$ in the same generation, and so we may write it as $p_k$. Moreover, for some value $k>2$, the probability $p_k$ is strictly positive. This follows by combining three observations:

\begin{enumerate}
\item
The probability that $v$ is the only descendant of an individual living $k$ generations ago is bounded below by $P(1)^k$, where $P(1)$ is the probability that any individual $x$ is an only child, i.e., that at least one of its parents had only one offspring.
\item
$P(1) \geq e^{-2}$ independently of $N$. This is because,  under  the random biparental genealogical  model,
$P(1)\geq (1-\frac{1}{N})^{2(N-1)}$, which decreases monotonically with $N$ to $e^{-2}$ as $N$ grows.
\item
The probability that no genetic material is inherited along a single lineage of length $k$ is strictly positive for a value $k>2$ and independent of $N$ (see e.g.  \cite{don}).  This probability depends on $k$,  on the recombination rate, and on the length of the genome. To establish the bound from  Proposition~\ref{pro-first}, however, we only use the fact that the probability is nonzero for some finite $k$ \cite{don}.   
\end{enumerate}

Thus for some value $r_k>0$ we have: $p_k(v)\geq r_k$, independent of $N$ and $v$.

Next, let $1<c'<c$.  Then from \cite{cha}, the following event $\A_N$ holds with probability at least $1-\epsilon(N)$, with $\lim_{N\rightarrow \infty} \epsilon(N)=0$:
\begin{itemize}
\item[$\A_N$:]  at generation $G=\lceil c' \log_2(N)\rceil$ before the present, there is at least one common ancestor of all individuals at the present.  
\end{itemize}

Consider the following process for generating a super-ghost, when $\A_N$ holds. 
Select one of these common ancestors at generation $G$, and call it $v$.   If $v$ has a $k$-critical ancestor, $w$, then select $w$ as a super-ghost and stop.  Otherwise, select
an ancestor of $v$ at generation $G+k$ before the present. This ancestor (call it $v'$) is also a common ancestor of every individual at the present. Whether $v'$ has a $k$-critical ancestor is independent of $\A_N$ and of whether $v$ has a $k$-critical ancestor, since all these depend only on the genealogies among non-overlapping generations. In this iterative way, we can try to identify a $k$-critical ancestor at every $k$ generations back in time from generation $G$. Let us attempt this $M$ times, back to generation $G+Mk$. Since we seek to exhibit super-ghosts within $c \cdot \log_2(N)$ generations, where $c> c'>1$, we can choose $M$ up to \begin{equation}M(N)= \lfloor (c-c') \log_2(N)/k \rfloor.
\label{MN}
\end{equation} This can be made arbitrarily large by increasing $N$. 

Thus, the  probability that the process described  fails to locate a super-ghost within $Mk+G$ generations conditional on $\A_N$ is bounded above by 
$(1-r_k)^M,$ and this can be made less than any $\delta>0$ by ensuring that $N$ is large enough that $M(N) >\frac{\log_2(1/\delta)}{-\log_2(1-r_k)}.$ Consequently,  the probability of the existence of a super-ghost conditional on $\A_N$ is at least $1-\delta$. Since $\lim_{N \rightarrow \infty} Pr(\A_N)=1$, and $\delta$ can be chosen arbitrarily small,  the probability of a super-ghost can be made as close to 1 as we like by selecting $N$ sufficiently large. This establishes the result.

The above argument can be extended to handle extensions of Chang's model to allow for population structure, as described in \cite{cha2}.  Note that we do not expect that our bounds are tight, which is why we have turned to simulations for estimates that take into account the different correlations among individuals. 

\bigskip

{\em Proof of Proposition~\ref{pro-second}}:   Label the positions along the genome of each individual $1,2, \ldots, B$.  For a given position $x$ let $M_T= M_T(x)$ be the number of individuals in generation $T$ before
the present for whom there is a path of genetic inheritance of position $x$ to at least one individual at the present. Under a Chang-type Markovian model of genealogical ancestry and
an associated Markovian process of recombination, the sequence $M_0 = N, M_1, M_2 \ldots$ is a finite state Markov chain with state space $1, \ldots, B$, and this chain has a strictly positive transition probability for the transition $i \rightarrow j$ precisely when $i\geq j \geq 1$. Moreover, $p_{ii}=1$ only for $i=1$. Thus $i=1$ is the unique absorbing state of this Markov process, and so:
\begin{equation}
\label{pone}
\lim_{T \rightarrow \infty}Pr(M_T(x)=1) =1.
\end{equation}

Let $\E_T(x)$ be the event that $M_T(x)=1$. Notice that the collection of events $\{\E_T(x): x \in \{1, \ldots, B\}\}$ are not independent (through shared genealogical ancestry, and also the resulting shared recombination history) however Eqn.~(\ref{pone}) implies that we nevertheless have:
\begin{equation}
\label{fix}
\lim_{T \rightarrow \infty}Pr(\bigcap_{x=1}^B \E_T(x)) =1,
\end{equation}
since $\Pr(\bigcap_{x=1}^B \E_T(x))  \geq 1- \sum_{x=1}^B (1-\Pr(\E_T(x))$, by Boole's inequality.
Now $\E_{N,T}:= \bigcap_{x=1}^B \E_T(x)$ is the event that each of the $B$ sites in the present population of size $N$ trace back to single ancestors $T$ generations ago.
Conditional on this event, these ancestral individuals need not all be different, but the number of them is at most $B$, and the remaining $N-B$ individuals contribute nothing genetically to the present population.  Thus, if we let 
$\G_{N,T}$ denote the event that a randomly selected ancestor $T$ generations in the past contributes nothing genetically to the present population of size $N$, then (for all $N,T$):
$$Pr(\G_{N,T}|\E_{N,T}) \geq (N-B)/N = 1-B/N,$$ and so:
\begin{equation}
\label{pipe}
\lim_{N \rightarrow \infty} \lim_{T \rightarrow \infty} Pr(\G_{N,T}|\E_{N,T})=1.
\end{equation}
Now,  $$q(N,T) = Pr(\G_{N,T})  \geq  Pr(\G_{N,T}|\E_{N,T}) Pr(\E_{N,T}),$$
and combining this with  Eqn. (\ref{pipe}) and $\lim_{T \rightarrow \infty} Pr(\E_{N,T}) =1$ (from Eqn. (\ref{fix})) we obtain Part (i).

For Part (ii), notice that $q'(N,T)$ is the probability that a randomly selected individual $T$ generations ago simultaneously satisfies two properties; namely (A) making no genetic  contribution to the present population, and (B) having every individual at the present as a genealogical descendant. Now, in the  limit as $T \rightarrow \infty$, 
the proportion of randomly selected individuals $T$ generations ago that satisfy property (A) is at least  $1-B/N$, with probability converging to 1 (from Eqn. (\ref{fix})) while the proportion of individuals that satisfy property (B) converges in probability to $1-\rho\approx 0.7968$. Consequently, in the limit as $T \rightarrow \infty$,  the probability that a randomly selected individual $T$ generations ago satisfies both properties ((A) and (B)) lies within a term of order $\frac{1}{N}$ from $1-\rho$.   
Finally, taking the outer limit, namely $N \rightarrow \infty$, we obtain  Part (ii). 
This completes the proof.

{\em Conversion of haploid to diploid values}

We wish to compare our diploid results with those of Baird et al. \cite{Baird}, who considered survival probability of  a single haploid genome (rather than a diploid genome), let $p_d$ be the haploid probability of leaving no genetic descendants for a single haploid genome, $p_n$ the haploid probability of leaving genealogical descendants but no genetic material (i.e., the probability of being a ghost), $p_g$ the haploid probability of being a ghost conditional on leaving genealogical descendants, and $\rho\simeq 0.2032$ the probability of leaving no pedigree descendants (haploid or diploid). We have  $p_n=(1-\rho) p_g$, and $p_d=\rho+p_n$. A diploid individual leaves a Poisson number of haploid genomes, each of which evolves independently in the infinite-population model of Baird et al. Let $p_D$, $p_N$, and $p_G$ be the corresponding probabilities for a diploid individual. For a constant population size, the mean of this Poisson distribution is 2,  and so $$p_D=\sum_{n \geq 0} \frac{e^{-2} 2^n}{n!} p_d^n=e^{-2+2p_d}. $$ 
Substituting $p_d= \rho + (1-\rho)p_g$ into this last equation, and noting that $p_N=p_D-\rho$  we obtain:
$$p_N = e^{-2(1-\rho)(1-p_g)} - \rho.$$

%
%

\end{document}